\documentclass{webofc}
\usepackage[varg]{txfonts}   
\usepackage[figurename=Fig.,labelfont=bf]{caption}
\usepackage{mathtools,braket}
\usepackage{soul}
\usepackage{lipsum}  
\usepackage{color}
\usepackage{amsmath,bm}
\usepackage{slashed}
\usepackage{float}
\usepackage{multirow}
\usepackage{xcolor}
\usepackage{physics}
\usepackage{multirow}
\usepackage{mathptmx}
\usepackage[colorlinks=true, pdfstartview=FitV, bookmarks=true, bookmarksnumbered=true, breaklinks]{hyperref}
\DeclareCaptionLabelSeparator{dot}{\textbf{. }}
\makeatletter
\renewcommand\@biblabel[1]{#1.}
\makeatother

\begin{document}
\title{The Properties of Radially Excited Charmonia\\  
in the Light Front Quark Model}
%
%

\author{\firstname{Muhammad Ridwan} \inst{1} \and
        \firstname{Ahmad Jafar Arifi}\inst{2,3}\thanks{Corresponding author: ahmad.arifi@riken.jp}\and
        \firstname{Terry Mart}\inst{1}
}

\institute{Departemen Fisika, FMIPA, Universitas Indonesia, Depok 16424, Indonesia 
\and
           Few-Body System in Physics Laboratory, RIKEN Nishina Center, Wako 351-0198, Japan 
\and
           Research Center for Nuclear Physics, Osaka University, Ibaraki, Osaka 567-0047, Japan
          }

\abstract{%
  Investigating the properties of excited charmonia is important to clarify its internal structure. In this paper, we present the mass spectra (MS) and decay constants (DC) for charmonia up to 3$S$ states calculated by means of the light-front quark model based on a variational approach. In particular, we consider the QCD-motivated effective Hamiltonian, which includes both confinement (linear and screened) and Coulomb-like potentials. Furthermore, since the existence of the nature of heavy quark symmetry, we treat hyperfine interactions perturbatively. We developed the harmonic oscillator expansion method to approximate the wave function (WF) for excited states. We found that the results of our theoretical calculations, using screened potentials rather than linear ones, are in good agreement with experimental data. By looking at the mass and decay constant result, we found that our result on the $\psi$(3S) state matched the properties of the $\psi$(4040) resonance.\\
  
}
\maketitle
\section{\label{Intro}Introduction}
Since 50 years ago, Quantum Chromodynamics (QCD)~\cite{Gross:2022hyw} has been understood as a fundamental theory of strong interaction between quarks and gluon with a non-perturbative nature in low-energy region and asymptotic freedom in high-energy limit.
With the emergence of a connection between QCD and constituent quark model (CQM), QCD becomes the most useful tool for describing hadrons, including baryons ($qqq$) and mesons ($q\bar{q}$).
However, quark models cannot provide a detailed explanation of exotic hadrons, such as the $X$, $Y$, and $Z$ states observed in experiments~\cite{Hosaka:2016pey}. 
Additionally, since the finding of the heavy charmonium state $J/\psi$ in 1974~\cite{Khare:1999zz}, experimentalists have also identified excited states of charmonium, particularly within the $\psi$ family~\cite{ParticleDataGroup:2022pth}, including $\psi(4040)$, $\psi(4160)$, and $\psi(4415)$. 
While numerous unknown hadrons have been confirmed in experiments, theorists are still working to develop suitable models that can explain the properties of hadrons, including some basic properties, namely mass spectra (MS) and decay constants (DC).

Since Dirac proposed the three relativistic dynamics in 1949~\cite{Dirac:1949cp}, light-front dynamics (LFD) has emerged as an effective tool for describing hadron structure.
In addition to employing a relativistic framework, LFD also offers distinctive features, including a rational energy-momentum relation and seven time-independent kinematic quantities.
Hence, calculating the properties of all hadrons becomes easier.
In particular, the light-front quark model (LFQM) offers a distinctive way of constructing hadron models by treating hadrons as relativistic bound states, making the model provide reliable predictions for hadron properties. 

There have been several efforts to determine the properties of mesons, including the calculation of MS and DC of the ground state~\cite{Choi:2015variational}, as well as investigation of the basis expansion effects of $\Phi(k) = \sum_{n=1}^2\phi^{HO}_{nS}(k)$ in the calculation up to the $2S$ state in LFQM~\cite{Arifi:2022pal}. The present work is mainly intended to calculate the properties of the $c\bar{c}$ charmonia, including MS and DC up to the $3S$ states. 
In particular, we investigate the effect of the inclusion of additional 3$S$ harmonic oscillator (HO) basis $\Phi(k) = \sum_{n=1}^3\phi^{\rm HO}_{nS}(k)$ in the radial part of the WF.

\section{\label{Model}Model Description}
In LFQM, our formalism utilizes the hadron state approximation through the leading-order Fock state expansion. 
The meson structure is considered at rest as an interacting bound system of effectively dressed quarks $q\bar{q}$, which satisfies the eigenvalue equation of the effective Hamiltonian
\begin{equation}
H_{q\bar{q}} \ket{\Psi_{q\bar{q}}} = M_{q\bar{q}} \ket{\Psi_{q\bar{q}}}
\end{equation}
where $M_{q\bar{q}}$ and $\Psi_{q\bar{q}}$ are the mass eigenvalue and eigenfunction, respectively. 
The Hamiltonian is given by
\begin{equation}\label{eq:2}
H_{q\bar{q}} = H_0 + V_{q\bar{q}} = \sqrt{m_q^2 + \pmb{ k}^2}  +  \sqrt{m_{\bar{q}}^2 + \pmb{ k}^2} + V_{q\bar{q}}, 
\end{equation}
where $H_0$ is the kinetic energy part of $q\bar{q}$ with three-momentum $\pmb{k}$. The effective potential $V_{q\bar{q}}$ is given by
\begin{equation}
V_{q\bar{q}} =  V_{\rm Conf} + V_{\rm Coul} + V_{\rm Hyp},
\end{equation}
where we use the linear confinement 
\begin{equation}
V_{\rm Conf} = a + br,
\end{equation}
where $a$ and $b$ are confinement parameters. The screened effect is given by
\begin{equation}
V_{\rm Conf} = a + \frac{b(1-{\rm e}^{-\mu r})}{\mu}.
\end{equation}
where $\mu$ is the screened parameter. The effective one-gluon exchanges for the $S$-wave mesons are assumed for the Coulomb and hyperfine interaction potentials, i.e., 
\begin{equation} 
V_{\rm Coul} = -\frac{4\alpha_s}{3r}, \hspace{0.5cm} {\rm and} \hspace{0.5cm} 
V_{\rm Hyp} = \frac{2 \expval{ \pmb{S}_q \cdot \pmb{S}_{\bar{q}} } }{3m_q m_{\bar{q}} } \nabla^2 V_{\text{Coul}}.
\end{equation}
The coupling constant $\alpha_s$ in this work is assumed to be a free parameter, whereas the values of  $\expval{ \pmb{S}_q \cdot \pmb{S}_{\bar{q}} }$ are $1/4$ ($-3/4$) for the vector (pseudoscalar) mesons, respectively. In addition, we consider the $\nabla^2 V_{\text{Coul.}}$ as the Dirac delta function $\delta^3(r)$.

The light-front wave function can be expressed as
\begin{equation}
\label{LFwavefunction}
\Psi^{M}_{q\bar{q}} = \Psi^{JJ_z}_{\lambda_q,\lambda_{\bar{q}}}(x,\pmb{k}_{\perp}) = \Phi(x,\pmb{k}_{\perp}) \mathcal{R}^{JJ_z}_{\lambda_q,\lambda_{\bar{q}}}(x,\pmb{k}_{\perp}),
\end{equation}
where $\Phi(x,\pmb{k}_{\perp})$ represents the radial wave function, and $\mathcal{R}^{JJ_z}_{\lambda_q,\lambda_{\bar{q}}}(x,\pmb{k}_{\perp})$ is the spin-orbit wave function derived from the Melosh transformation.
The covariant forms of $\mathcal{R}^{JJ_z}_{\lambda_q\lambda_{\bar{q}}}$ are given by
\begin{equation}\label{eq:6}
	\mathcal{R}^{JJ_z}_{\lambda_q\lambda_{\bar{q}}} =  \frac{1}{\sqrt{2} \tilde{M}_0} 
	\bar{u}_{\lambda_q}^{}(p_q) \Gamma_{P(V)} v_{\lambda_{\bar{q}}}^{}(p_{\bar{q}}), 
\end{equation}
 where $\Gamma_{P(V)}$ is the vertex for pseudoscalar(vector) meson,
\begin{equation}
\Gamma_P = \gamma_5,  \hspace{0.5cm} {\rm and} \hspace{0.5cm} 
\Gamma_V = -\slashed{\epsilon}(J_z) + \frac{\epsilon \cdot (p_q-p_{\bar{q}})}{M_0 + m_q + m_{\bar{q}}},
\end{equation}
with $\tilde{M}_0 \equiv \sqrt{M_0^2 - (m_q -m_{\bar{q}})^2}$. 
The polarization vectors $\epsilon^\mu(J_z)=(\epsilon^+, \epsilon^-,\pmb{\epsilon}_{\perp})$  
are given by
\begin{equation}\label{eq:7}
\epsilon^\mu(\pm 1) = \left( 0, \frac{2}{P^+} \bm{\epsilon}_\perp(\pm) \cdot \pmb{ P}_\perp, \pmb{\epsilon}_\perp(\pm)\right), \hspace{0.5cm} {\rm and} \hspace{0.5cm} 
\epsilon^\mu(0) = \frac{1}{M_0}\left(P^+, \frac{-M^2_0 + \pmb{P}^2_\perp}{P^+}, \pmb{P}_\perp\right),
\end{equation}
where $\pmb{\epsilon}_\perp(\pm 1) = \mp\left( 1, \pm i \right)/\sqrt{2}$.
Since we only consider the radial part, the spin-orbit part satisfies the unitary condition, i.e.,
$\Braket{ \mathcal{R}^{JJ_z}_{\lambda_q\lambda_{\bar{q}}} | \mathcal{R}^{JJ_z}_{\lambda_q\lambda_{\bar{q}}} } = 1$.
The light-front wave function is written in terms of the invariant Lorentz variable $x_i = p^+_i / P^+$, $\pmb{k}_{\perp i} = \pmb{p}_{\perp i} - x_i \pmb{p}_{\perp}$, helicity $\lambda_i$, $P^{\mu} = (P^+, P^-, \pmb{p}_{\perp})$ is the meson's momentum and $p^{\mu}$ is the $i$-th ($i$=1,2) constituent quark's momentum, and define $x \equiv x_1$ with $\pmb{k}_\perp \equiv \pmb{k}_{\perp 1}$. Furthermore, the three-momentum $\pmb{k} = (k_z, \pmb{k}_\perp)$ can be described as $\pmb{k} = (x, \pmb{k}_\perp)$ through the relationship,
\begin{equation}
k_z = \left( x- \frac12 \right) M_0 + \frac{m^2_{\bar q}-m^2_q}{2M_0}, 
\end{equation}
where
\begin{equation}\label{eq:4}
M_0^2 = \frac{\pmb{k}_{\bot}^2 + m_q^2}{x}  + \frac{\pmb{k}_{\bot}^2 + m_{\bar{q}}^2}{1-x}
\end{equation}
the boost-invariant meson mass squared is introduced in Eq. (\ref{eq:4}). Note that the normalization is taken into account on our radial part following the variable transformation $\{k_z, \pmb{k}_\bot \} \to \{x, \pmb{k}_\bot \}$ emerges and influence the Jacobian factor $\frac{\partial k_z}{\partial x}$
\begin{equation}
\frac{\partial k_z}{\partial x} = \frac{M_0}{4x(1-x)} \left[ 1 - \frac{ (m_q^2 - m_{\bar{q}}^2)^2}{M_0^4} \right],
\end{equation}
For the ground state, we use the harmonic oscillator expansion method (HOEM) in our radial wave function as
\begin{eqnarray}\label{eq:9}
	\phi_{1S}^{H0} (\pmb{k}) &=& \frac{1}{ \pi^{3/4}\beta^{3/2}} e^{-\pmb{k}^2/ 2\beta^2},\\
	\phi_{2S}^{H0} (\pmb{k}) &=& \frac{(2k^2 -3\beta^2)}{\sqrt{6} \pi^{3/4}\beta^{7/2}} e^{-\pmb{k}^2/ 2\beta^2},\\
 	\phi_{3S}^{H0} (\pmb{k}) &=& \frac{(15\beta^4 -20\beta^2k^2 + 4k^4)}{2\sqrt{30} \pi^{3/4}\beta^{11/2}}  e^{-\pmb{k}^2/ 2\beta^2},
\end{eqnarray}
with $k=|\pmb{k}|$ and
\begin{equation}
    \phi_{nS}^{H0}(x,\pmb{k}_\bot) = \sqrt{2(2\pi)^3} \sqrt{\frac{\partial k_z}{\partial x}} \phi_{nS}^{H0}(\pmb{k}),
\end{equation}
whereas $\beta$ is the parameter that is inversely proportional to the width of the WF. In this calculation, $\beta$ is considered as the variational parameter. The WF $\phi_{nS}$ include with the Jacobian factor $\partial k_z/\partial x$ satisfy the normalization:
\begin{equation}\label{eq:10}
 \int \frac{\dd x \dd^2 \pmb{k}_\bot}{2(2\pi)^3}  |\phi_{nS}^{H0}(x, \bm{k}_\bot)|^2 =1.
\end{equation}
Furthermore, we expand the basis as a linear combination of the $1S$, $2S$, and $3S$ HO bases. 
It can be written in a matrix form
\begin{eqnarray}
\begin{pmatrix}
\Phi_{1S}  \\
\Phi_{2S}  \\
\Phi_{3S}  
\end{pmatrix}
&=&
\begin{pmatrix}
c_{12}c_{13} & s_{12}s_{13} & s_{13} \\
-s_{12}c_{23} - c_{12}s_{23}s_{13} & c_{12}c_{13} - s_{12}s_{23}s_{13} & s_{23}c_{13} \\
s_{12}c_{23} - c_{12}s_{23}s_{13} & -c_{12}c_{23} - s_{12}c_{23}s_{13} & c_{23}c_{13} 
\end{pmatrix}
\begin{pmatrix}
\phi_{1S}^{\rm HO}  \\
\phi_{2S}^{\rm HO} \\
\phi_{3S}^{\rm HO}  
\end{pmatrix} \nonumber \\
&=&
\begin{pmatrix}
c_1^{1S} & c_2^{1S} & c_3^{1S} \\
c_1^{2S} & c_2^{2S} & c_3^{2S} \\
c_1^{3S} & c_2^{3S} & c_3^{3S} 
\end{pmatrix} 
\begin{pmatrix}
\phi_{1S}^{\rm HO} \\
\phi_{2S}^{\rm HO} \\
\phi_{3S}^{\rm HO}
\end{pmatrix} \, ,
\end{eqnarray}
with $c_{ij}(s_{ij}) = \cos\theta_{ij}(\sin\theta_{ij})$. Note that the matrix form in $R_{M}$ we used is quite similar to the CKM matrix~\cite{Chau:1984fp}, but we ignore the CP violation terms.
In our LFQM analysis, the heavy mesons (in this case, charmonia) have a number of parameters, i.e., the constituent quark masses ($m_c$), the potential parameters $(a, b,\alpha_s, \mu)$, the HO parameter $\beta_{cc}$ for both pseudoscalar and vector charmonia. We begin with determining the parameters' values by fitting the MS based on the variational calculation
\begin{equation}\label{eq:11}
\frac{\partial \bra{\Psi_{q\bar q}}  H_0  \ket{\Psi_{q\bar q}} }{\partial \beta} = 0,
\end{equation}
and then computing the other meson properties such as decay constant.
We note that the hyperfine interaction is treated perturbatively.

\subsection{Mass Spectra}

The mass formula of 1S, 2S, and 3S state charmonia with three HO bases is expressed as
\begin{equation}
\label{mass_spectra}
    M_{q\Bar{q}} = \expval{ H_0} + \expval{V_{\rm Conf}} + \expval{V_{\rm Coul}} + \expval{V_{\rm Hyp}},
\end{equation}
where the formula is derived by taking the expectation values of the effective Hamiltonian
\begin{equation}
    \bra{\Psi_{q\Bar{q}}} (H_0 + V_{q\Bar{q}})\ket{\Psi_{q\Bar{q}}}.
\end{equation}
The contribution from the kinetic term is given by
\begin{eqnarray}
   \expval{H_0} &=& \frac{\beta}{120\sqrt{\pi}}  \sum_{j=1,2} \biggl[  ( 120c_1^2 - 120\sqrt{6} c_1 c_2 + 180 c_2^2 + 60 \sqrt{30} c_1 c_3 - 180\sqrt{5} c_2 c_3  \nonumber \\
   && + 225 c_3^2 + 40 c_2^2 + 8\sqrt{30} c_1 c_3 -104\sqrt{5} c_2 c_3 +  260 c_3^2 z_i^2) z_i {\rm e}^{z_i/2} K_1 \left[ \frac{z_i}{2}\right] \nonumber\\
   &&- 4(10 c_2^2 -26\sqrt{5} c_2 c_3 + 2 \sqrt{30}c_1 c_3 + 65 c_3^2) z_i^2 (z_i^2 -3) {\rm e}^{z_i/2} K_2 \left[ \frac{z_i}{2}\right] \nonumber\\
   &&+ 15\sqrt{\pi} \biggl( 4(-6 c_2^2 + 9\sqrt{5} c_2 c_3 - 15 c_3^2  + 2 \sqrt{6}( c_1 c_2 - \sqrt{5} c_1 c_3))  \nonumber \\
   && \times U(-\tfrac{1}{2},-2,z_i) + 28 (\sqrt{5} c_2 c_3 - 5 c_3^2) U(-\tfrac{1}{2},-4,z_i)+ 63 c_3^2 U(-\tfrac{1}{2},-5,z_i) \biggr) \biggl], \nonumber\\  
\end{eqnarray}
where $z_i = m_i^2/\beta^2$, $K_n(x)$ is the modified Bessel function of second kind of order \textit{n}, and $U(a,b,z)$ is the Tricomi's (confluent hypergeometric) function. The contribution from the linear confinement $(a+br)$  is given by
\begin{eqnarray}
   \expval{V^{\rm Lin}_{\rm Conf}} &=& a  + \frac{b}{\beta \sqrt{\pi}}\biggl(2c_1^2 + 3c_2^2 + \frac{15}{4}c_3^2 - 2\sqrt{\frac{2}{3}}c_1 c_2 - \sqrt{\frac{2}{15}}c_1 c_3  - \sqrt{5}c_2 c_3 \biggr),\quad \quad 
\end{eqnarray}
and those of screened confinement $( a + {b(1-{\rm e}^{-\mu r})}/{\mu})$ is given by
\begin{eqnarray}
\label{eqs:28}
  \expval{V^{\rm Scr}_{\rm Conf}} &=& a + \frac{b}{\mu} + \frac{b}{3840\sqrt{\pi}\beta^{10}\mu} \biggl[ 2\beta \mu \bigl( 32(60 c_1^2 - 4\sqrt{6}(10 c_1 c_2 + \sqrt{5} c_1 c_3) \nonumber \\
  &&+ 15(8c_2^2 -4\sqrt{5}c_2 c_3 + 11 c_3^2)) \beta^8 + 32(40 c_2^2 -48\sqrt{5}c_2 c_3 + 115 c_3^2 \nonumber \\
  && + 2\sqrt{6}(-5 c_1 c_2 + 2\sqrt{5} c_1 c_3) \beta^6\mu^2 + 8 (10c_2^2 -28\sqrt{5} c_2 c_3\nonumber \\
  && + 2\sqrt{30} c_1 c_3 + 89 c_3^2 ) \beta^4\mu^4 + -8(\sqrt{5} c_2 c_3 - 6 c_3^2 ) \beta^2 \mu^6 + c_3^2 \mu^8 \bigr) \nonumber \\
  &&- \sqrt{\pi}{\rm e}^{\mu^2/4\beta^2} ( 3840 \beta^{10} + 1920( c_1^2 + 3c_2^2 + 5 c_3^2 - \sqrt{6} c_1 c_2 \nonumber \\
  && - 2\sqrt{5} c_2 c_3) \beta^8 \mu^2 + 160( 9 c_2^2 + 30 c_3^2 + \sqrt{30} c_1 c_3 - 2\sqrt{6}c_1 c_2\nonumber \\
  &&  - 12\sqrt{5} c_2 c_3  ) \beta^6 \mu^4 + 16( 5 c_2^2 + 50 c_3^2 - 15\sqrt{5} c_2 c_3 + \sqrt{30} c_1 c_3 ) \beta^4 \mu^6  \nonumber \\
  && + 50 c_3^2 - 8\sqrt{5}c_2 c_3 \beta^2 \mu^8 + c_3^2 \mu^{10} ) {\rm erfc}\left[\frac{\mu}{2\beta}\right]\biggl], 
\end{eqnarray}
where the error function ${\rm erfc}(z) = 1 - {\rm erf}(z)$. The Coulomb-like and Hyperfine potentials are given by
\begin{eqnarray}
  \expval{V_{\rm Coul}} &=& - \frac{\beta \alpha_s}{45\sqrt{\pi}} \biggl( 120c_1^2 +100c_2^2 + 89 c_3^2 + 40\sqrt{6} c_1 c_2 + 12\sqrt{30} c_1 c_3 \nonumber \\
  &&+ 44\sqrt{5}c_2 c_3 \biggr), \\ 
  \expval{V_{\rm Hyp}} &=& \frac{\expval{\pmb{S}_q\cdot\pmb{S}_{\Bar{q}}}\beta^3 \alpha_s}{3 m_c m_{\Bar{c}}\sqrt{\pi}} \biggl( \frac{32}{3}c_1^2 + 16 c_2^2 + 20 c_3^2 + 32\sqrt{\frac{2}{3}} c_1 c_2 + 16\sqrt{\frac{10}{3}} c_1 c_3 \nonumber \\
  && + 16 \sqrt{5} c_2 c_3\biggr),
\end{eqnarray}
respectively.
Note that the meson's mass is able to be calculated by replacing $c_i$ in the mass formula with $c_i^{1S}, c_i^{2S},$ or $c_i^{3S}$ for $1S$, $2S$, and $3S$ state, respectively.

\subsection{Decay Constant}
All pseudoscalar(vector) mesons $P$($V$) with four-momentum $P^\mu$ and mass $M_P$ ($M_V$) have decay constants, namely $f_P$ and $f_V$, which are defined in the matrix transition form as 
\begin{eqnarray}\label{eq:17}
	\bra{0} \bar{q} \gamma^\mu \gamma_5 q \ket{P} &=& i f_P P^\mu, \nonumber\\
	\bra{0} \bar{q} \gamma^\mu q \ket{V(P,\lambda)} &=& f_V M_V \epsilon^\mu(\lambda),
\end{eqnarray}
where $\epsilon^\mu (\lambda)$ is the polarization vector of a vector meson. Solving the left and right-hand side of Eq. (\ref{eq:17}) simultaneously, we obtain the explicit form of $f_{P(V)}$
\begin{eqnarray}\label{eq:18}
	f_P &=& 2\sqrt{6} \int \frac{\dd x \dd^2 \pmb{k}_\bot}{2(2\pi)^3}  
	\frac{ {\Phi}(x, \pmb{k}_\bot) }{\sqrt{\mathcal{A}^2 + \mathbf{k}_\bot^2}} ~\mathcal{A},\\
	f_V &=& 2\sqrt{6} \int\frac{\dd x \dd^2 \pmb{k}_\bot}{2(2\pi)^3}   
\frac{ {\Phi}(x, \pmb{k}_\bot) }{\sqrt{\mathcal{A}^2 + \pmb{k}_\bot^2}} 
\left[\mathcal{A} + \frac{2 \pmb{k}_\bot^2}{D_{0}}\right], 
\end{eqnarray}
respectively, where $\mathcal{A} =  (1-x)  m_q+  x m_{\bar{q}}$ and and $D_{0}= M_0 + m_q + m_{\bar{q}}$.

\section{Result and Discussion}

Before calculating the mass spectra and decay constants, it is necessary to determine seven parameters through the variational principle applied to $\Phi_{1S}$ in the mass formula. 
The values of these parameters are listed in Table~\ref{tab:parameter}. 
Here, we only fix the widely-known string tension $b=0.18$ GeV$^2$. 
Additionally, we assume $\theta_{13}=\theta_{23}$ to simplify the parameterization that reasonably describes the experimental data.
Initially, we considered $\theta_{12}=\theta_{13}=\theta_{23}$ and found that the predicted result deviates considerably from the data. 
Because of that, we assume two mixing angles for the basis expansion coefficients in this work.   
We obtain the $\theta_{12} = 12.12 ^{\circ}$ and $\theta_{13} = 8.44 ^{\circ}$ where $\theta_{12} > \theta_{13}$. As for comparison, the angle value $\theta_{12}$ is quite similar to Ref.~\cite{Arifi:2022pal}, but they consider only two HO bases such that ($\theta_{13} = \theta_{23}=0$) and use it to calculate observables for the $1S$ and $2S$ states. 

\begin{table}[h]
	\centering	
	\caption{Model parameters that include constituent quark mass $m$, potential parameters $(a$ and $b)$ as well as screening parameter $\mu$ and variational parameters $\beta$ (GeV).}
	\label{tab:parameter}
	\begin{tabular}{cccccccc}
		\hline
		$\theta_{12}$ & $\theta_{13}=\theta_{23}$ & $m_c$ & $a$ & $b$ & $\mu$ & $\alpha_s$ & $\beta_{c\bar{c}}$ \\ 
		\hline 
		$12.12$ & $8.44$ & $1.61$ & $-0.41$ & 0.18 & $0.027$ & $0.402$ & $0.5417$ \\
		\hline
	\end{tabular}
	\label{parameters}
\end{table}

In the case of the quark mass, our obtained value of 1.61 GeV is quite similar to the value in the Godfrey-Isgur (GI) model~\cite{Godfrey:1985xj}.
However, our results for the potential parameters, particularly the strong coupling constant $\alpha_s$, differ somewhat from Arifi's work. 
Moreover, our value for the screening parameter $\mu$ is a bit different from that in Gao's work~\cite{Gao:2022Screened}, where they obtained $\mu = 0.045$ GeV$^2$.
As for the variational parameter $\beta$ on the HO expansion, 
we obtain $\beta_{c\bar{c}} = 0.5417$ GeV. 
We note that other parameters such as the mixing angle and potential parameter will influence the value of $\beta$. 
For example, Arifi et al., obtained $\beta_{c\bar{c}} = 0.592$ GeV where they only apply linear potential and $\theta = 12^{\circ}$. 
This is quite essential since the $\beta$ parameter determines the the wave function characteristics and is always inversely proportional to the width of the wave function~\cite{Wang:2021hho}. 
As a result, it indirectly affects the determination of MS and DC.

\begin{table}[h]
	\centering
	\caption{Mass spectra [MeV] with linear and screened potentials. We also show other model predictions for comparison.}
	\label{masslist}
	\begin{tabular}{c|cccccc}
		\hline
		\hline
		State &	$M_{\rm lin.}$ & $M_{\rm scr.}$ & Exp.~\cite{ParticleDataGroup:2022pth} & GI~\cite{Godfrey:1985xj} & RQM~\cite{Ebert:2003RQM} & NRQCD~\cite{Soni:2017wvy} \\ 
		\hline  
		$\eta_c(1S)$ & 3010.8 & 3002.3 &  2983.9 $\pm$ 0.4 & 2970 & 2979 & 2989 \\
		$\eta_c(2S)$ & 3641.3 &  3614.5 & 3637.5 $\pm$ 1.1 & 3620 & 3588 & 3602  \\
		$\eta_c(3S)$ & 4076.3 &  4028.1 & \dots & 4060 & 3991 & 4058  \\
		\hline
		$J/\psi(1S)$ & 3110.8 & 3102.4 & 3098.9 $\pm$ 0.01 & 3100 & 3096 & 3094  \\
		$\psi(2S)$ & 3706.6 & 3679.8 & 3686.1$\pm$ 0.06 & 3680 & 3686 & 3681  \\   
		$\psi(3S)$ & 4127.3 & 4079.1 & 4039 $\pm$ 1 & 4100 & 4088 & 4129  \\
		\hline
		\hline 
	\end{tabular}
\end{table}

First, let us discuss the obtained masses of the charmonia.
Here, we use two confinement potentials separately to study its effect. 
While $M_{\rm lin.}$ denotes the mass spectra using linear potential, $M_{\rm scr.}$ denotes the mass spectra using screened potential. 
We present our prediction along with the experimental value from PDG as well as the results from other theoretical models in Table~\ref{masslist}. 
We can clearly see that $M_{\rm scr.}$ for all charmonium states are lower than those in $M_{\rm lin.}$ and the discrepancy increases as the radial excitation level increases. 
It happens because of the exponential term in the screened potential, 
which leads to a decrease in mass as the distance increases. 
We note that the excited state wave function is more extended.
It is worth noting that the inclusion of the mixing angle in our calculation, whether for $M_{\rm lin.}$ or $M_{\rm scr.}$, effectively reduces the predicted mass. 
A similar observation can be found in Arifi's work~\cite{Arifi:2022pal}. 
Besides our predictions and experiments, results from the GI model~\cite{Godfrey:1985xj} are quite similar to the experiment, although there is a considerable difference on $\psi (3S)$.
For instance, some of the states on charmonia that employ the relativistic effect on the Relativized Quark Model (RQM)~\cite{Ebert:2003RQM} model will reduce the mass result compared to the Non-Relativistic Quantum Chromodynamics (NRQCD)~\cite{Soni:2017wvy} model although this model also uses the linear potential. 
Interestingly, our prediction and other theoretical models on the $\psi (3S)$ state lead to the $\psi (4040)$ instead of $\psi (4160)$ and $\psi (4415)$ as well.
A review from Ref.~\cite{Mo:2010bw} suggests that $\psi (4040)$ corresponds to the $\psi (3S)$ and $\psi (4415) = \psi (5S)$, while the assignment of $\psi (4160)$ to a specific state has not yet been determined.

\begin{table}[h]
	\centering
	\caption{Predicted decay constant [MeV] for charmonia.}
	\label{DClist1}
	\begin{tabular}{c|cc|ccc}
	\hline\hline
	State & $f_{\rm Theo.}$ & Exp.~\cite{ParticleDataGroup:2022pth} & Lattice~\cite{Davies:2010ip,Davies:2012ogo} & Sum Rules~\cite{Becirevic:2016rfu} & LFQM (CJ)~\cite{Choi:2009Semileptonic}  \\
	\hline
	$\eta_c$ ($1S$) & 349.1 & 335(75) & 395(2.4) & 387(7) & 326 \\
	$J/\psi$ ($1S$) & 396.7 & 407(5) & 405(6) & 418(9) & 360 \\
	\hline
                 &       &        & BLFQ~\cite{Li:2017mlw} & RQM~\cite{Azhothkaran:2020ipl} & RQM2~\cite{Lakhina:2006vg} \\
        \hline
        $\eta_c$ ($2S$) & 224 & \dots & 299(68) & 173 & 240  \\
	$\psi$ ($2S$) & 285.6 & 294(5) & 312(73) & 261 & 293  \\
        \hline
                 &       &        & NRQM~\cite{Soni:2017wvy} & RQM~\cite{Azhothkaran:2020ipl} & RQM2~\cite{Lakhina:2006vg} \\
        \hline
        $\eta_c$ ($3S$) & 163.9 & \dots & 249 & 127 & 193 \\
	$\psi$ ($3S$) & 230.2 & 238(5) & 230 & 206 & 258 \\
        \hline \hline
	\end{tabular}
\end{table}

Here, we have also calculated the decay constants for the charmonium states. Our predictions, along with a comparison to the experimental values and the results from other theoretical models, are presented in Table~\ref{DClist1}. 
To determine the experimental value of the decay constant for each charmonium state, we extract it using the leptonic partial width $\Gamma_{e^- e^+}$ for vector charmonium and the diphoton partial width $\Gamma_{\gamma\gamma}$ for pseudoscalar charmonium. 
It can be observed that the decay constant decreases as the radial excitation level increases, establishing the hierarchy $f(1S) > f(2S) > f(3S)$, where the ground state has a larger decay constant than the excited states. 
Unfortunately, experimental results for $\eta_c (2S)$ and $\eta_c (3S)$ cannot be found because there is no experimental data available for determining the partial width required to calculate the decay constant in the PDG. 

In general, our results for the decay constant are in good agreement with the experimental data, 
especially for $\psi$ mesons. 
However, for the $1S$ state, we obtained different results compared to other LFQM by the Choi-Ji (CJ) scheme~\cite{Choi:2009Semileptonic}.
This difference may be attributed to their focus being primarily on the ground state mesons with various quark flavors and variations in parameter determination during the variational analysis. 
We found that the results for $\eta_c (1S)$ are consistent with the experimental data but deviate from the lattice data~\cite{Davies:2010ip,Davies:2012ogo} and sum rules~\cite{Becirevic:2016rfu}.
For the $J/\psi (1S)$, our prediction is comparable to the experimental data and other calculations.
In a similar framework of light-front dynamics, basis light-front quantization (BLFQ)~\cite{Li:2017mlw} produces a higher result than ours for the $2S$ states, 
particularly showing a moderate difference between BLFQ and experimental result for $\psi(2S)$.
For the $3S$ state, our predictions for $\psi(3S)$ are generally consistent with the experimental data and other calculations. 
However, our result on $\eta_c$ has a large discrepancy compared to either RQM~\cite{Azhothkaran:2020ipl} or NRQM~\cite{Soni:2017wvy}, but not quite differ to RQM2~\cite{Lakhina:2006vg}. 
This discrepancy arises from differences in the treatment of certain parameters. 
Specifically, Lakhina's work~\cite{Lakhina:2006vg} defines the running constant $\alpha_s$ with a logarithmic scale applied to their denominator, whereas Azhothkaran~\cite{Azhothkaran:2020ipl} and Soni's work~\cite{Soni:2017wvy} takes into account QCD corrections, including $\delta_P$ (for pseudoscalar meson) and $\delta_V$ (for vector meson), in their decay constant calculation. 
Consequently, our result appears to be in better agreement with Lakhina's findings than with the others.

\section{Conclusion and Outlook}
We have analyzed the properties of charmonia ($\eta_c, J/\psi$) from the ground state ($1S$) to the second radial excitation ($3S$) using the light-front quark model. 
In this study, we calculated their mass spectra along with decay constants by employing the harmonic oscillator expansion method.
Furthermore, we expand the basis as a linear combination of $\phi_{1S}, \phi_{2S}$, and $\phi_{3S}$. To obtain the parameters $\beta_{cc}, a, b,\alpha_s,\mu$, and $\theta_{ij}$ 
we carried out the variational analysis using the mass formula we have derived.

For our mass spectra calculation, we consider the QCD-motivated effective Hamiltonian with several potentials such as confinement (linear and screened), Coulomb-like, and hyperfine term. 
In this work, we consider the hyperfine potential perturbatively since its contribution is rather small.
Our predictions for mass spectra up to $3S$ states with the screened potential are consistent with the experimental data and other theoretical models. 
Also, our predictions for decay constants are in good agreement with experimental data and other theoretical models.
We note that $f_{P(V)}^{exp}$ are extracted from $\Gamma_{e^-e^+}$ ($\Gamma_{\gamma\gamma}$) for vector (pseudoscalar) mesons. 
Interestingly, we find our model can reproduce the hierarchy of decay constant  $f_{1S} > f_{2S} > f_{3S}$ which is mainly due to the increasing number of the basis functions. If we use a single HO basis function, the hierarchy of the decay constant will be opposite~\cite{Arifi:2022pal}.

For future work, we would like to consider other mesons that contain bottom quarks such as $B_c^{(*)}$ mesons and bottomonia ($\eta_b, \Upsilon$). Computing other properties such as distribution amplitudes~\cite{Arifi:2023uqc} and radiative decays\cite{Choi:2007se} with the obtained light-front wave function is also certainly of interest.
We would like also to apply the smeared function~\cite{Hong:2022sht} instead of the Dirac delta function in the spin-spin interaction.
It is important to note that the Harmonic Oscillator (HO) basis function is assumed for the wave function in the present work, but one should study different basis functions further to solve the Hamiltonian that can lead to more accurate predictions.

\section*{Acknowledgements}
M.R. thanks the ISCPMS 2023 organizer for providing the oral presentation for his work. The work of A.J.A. is supported by the RIKEN special postdoctoral researcher (SPDR) program. The work of T.M. is supported by the PUTI Q2 grant from Universitas Indonesia (UI), under contract No. NKB-663/UN2.RST/HKP.05.00/2022. The hospitality of the master program in physics at UI is also gratefully acknowledged.

\end{document}